\documentstyle[12pt]{article}
\newcommand{\be}{\begin{equation}}
\newcommand{\ee}{\end{equation}}
\newcommand{\bea}{\begin{eqnarray}}
\newcommand{\eea}{\end{eqnarray}}
\newcommand{\nn}{\nonumber \\}
\newcommand{\p}{\partial}

\begin{document}

\begin{titlepage}

\vspace*{-2cm}

\hspace*{\fill} CGPG-96/11-1 \\
%\hspace*{\fill} ULB--TH--95/(??) \\
\vspace{.5cm}

\begin{centering}

{\huge Geometrical representation of Euclidean
general relativity in the canonical formalism}

\vspace{.5cm}
{\large Glenn Barnich$^{*,1}$ and
Viqar Husain$^{1,2}$}\\
\vspace{1cm}
Center for Gravitational Physics and Geometry,
The Pennsylvania State University,
104 Davey Laboratory, University Park, PA 16802-6300$^{1}$, \\
and Department of Mathematics and Statistics, University of
New Brunswick, Fredericton, N.B. E3B 5A3, Canada$^2$.
\vspace{.5cm}

\begin{abstract}
We give an $SU(2)$ covariant representation of the constraints of
Euclidean general relativity in the Ashtekar variables. The guiding
principle is the use of triads to transform all free spatial indices
into $SU(2)$ indices. A central role is played by a special covariant
derivative. The Gauss, diffeomorphism and Hamiltonian constraints
become purely algebraic restrictions on the curvature and the torsion
associated with this connection. We introduce coordinates on the
jet space of the dynamical fields which cleanly separate the constraint
and gauge directions from the true physical directions. This leads to
a classification of all local diffeomorphism and Gauss invariant
charges.

\end{abstract}

\vspace{1.5cm}

{\footnotesize \hspace{-0.6cm}($^*$)Aspirant au Fonds National de la
Recherche Scientifique (Belgium). On leave of absence from Universite
Libre de Bruxelles.}

\end{centering}

\end{titlepage}

\pagebreak

\def\qed{\hbox{${\vcenter{\vbox{
    \hrule height 0.4pt\hbox{\vrule width 0.4pt height 6pt
   \kern5pt\vrule width 0.4pt}\hrule height 0.4pt}}}$}}
\newtheorem{theorem}{Theorem}
\newtheorem{lemma}{Lemma}
\newtheorem{definition}{Definition}
\newtheorem{corollary}{Corollary}
\newcommand{\proof}[1]{{\bf Proof.} #1~$\qed$.}
\renewcommand{\theequation}{\thesection.\arabic{equation}}
\renewcommand{\thetheorem}{\thesection.\arabic{theorem}}
\renewcommand{\thelemma}{\thesection.\arabic{lemma}}
\renewcommand{\thecorollary}{\thesection.\arabic{corollary}}
\renewcommand{\thedefinition}{\thesection.\arabic{definition}}

\section{Introduction}

The recent progress in non-perturbative quantum gravity using
Ashtekar's formulation of general relativity is due, in part, to
the application to gravity of techniques used for studying
Yang-Mills theory non-perturbatively \cite{Ash}. The constraints
of Euclidean or Lorenzian general relativity are appealing in this
formulation because they are polynomials of order at most four in the
basic variables. An example of the progress is a complete
non-perturbative
quantization \cite{ALMMT} of the Husain-Kucha\v r model. This model
consists of
a four-dimensional generally covariant
$SU(2)$
gauge theory, which happens to be perturbatively non-renormalizable,
and has a phase space like that of general relativity, except that
the
Hamiltonian constraint vanishes identically \cite{HK}. More recently, a
possibly complete non-perturbative quantization of general relativity
has
been given by Thiemann \cite{TT}.

An important aspect of the approach
is the use of the $SU(2)$ invariant Wilson loops as elementary
classical variables of the theory. There is a non-countable number of
such elementary variables, since they are labelled by the
(inequivalent) loops around which the holonomies are evaluated. In
order to control this configuration space,
it is given as the projective limit of finite dimensional
spaces associated with a finite number of inequivalent loops.

The aim of this paper is a better understanding of the reduced phase
space of Euclidean general relativity in the Ashtekar variables.
As a first step in this
direction, we give a complete classification of all $SU(2)$ and
diffeomorphism invariant local quantities. At the same time, this
corresponds to a complete classification of the local conservation
laws of the Husain-Kucha\v r model. The characterization ``local''
comes from the fact that we work in the context of jet-spaces, which
provide an appealing (countable) projective family for analytic sections.

During the analysis, we are naturally led to a special covariant
derivative, given by the $SU(2)$ covariant derivative, where the
spacetime indices are converted into $su(2)$ indices using the inverse
triads. A first result, of considerable interest in itself, is that
the constraints can be rewritten in a purely geometrical way.
The diffeomorphism constraint corresponds to the vanishing of the trace
of the
curvature of this covariant derivative,
and the Gauss constraint corresponds to
the vanishing
of the trace of its torsion. The Hamiltonian constraint corresponds to
an additional
algebraic restriction on the curvature.

The paper is organized as follows. In the next section, we fix the
notations
and define the models. We then introduce the special covariant
derivative, and
give the covariant representation of the constraints and their
algebra.
The following two sections are devoted to explaining how this
representation may
be arrived at in a constructive way.

In the third section we give some
ideas about jet-spaces as applied to $SU(2)$ Yang-Mills theory.
A detailed analysis of the orbit space and Wilson loops in this
context serves both as a warm up before discussing the more
complicated case of diffeomorphism invariance, and as a possible bridge
for comparison with the quantization using Wilson loops as fundamental
variables.
We first present a
change of coordinates which allows the separation of coordinates purely
along the gauge orbits from coordinates containing the gauge invariant
information. We then show how the local information contained in the
Wilson loops can be expressed in terms of these latter
coordinates.

In the fourth section, we apply these ideas to the
diffeomorphism
and  Gauss constraints of Euclidean general relativity in Ashtekar's
variables
to obtain the geometrical representation of the constraint surface.

The classification of all the local conservation laws of the
Husain-Kucha\v r model \cite{HK}, or equivalently, all local
diffeomorphism
and $SU(2)$ invariant quantities is done in section 5 and
a corresponding appendix, containing
the local BRST cohomology of the model.

Finally, we consider models with the Hamiltonian constraint function as
the integrand of an action. In three dimensions such an action is
topological
in the sense that its field equations require both the curvature and the
torsion of the covariant derivative to vanish. In four dimensions, the
covariant field equations give an identically vanishing Hamiltonian
constraint.

\section{Geometrical representation of constraints}
\setcounter{equation}{0}

Let us first fix the conventions. The indices $i,j,k,\dots$ denote the
$SU(2)$ indices which are raised and lowered with the Euclidean metric,
while the indices $a,b,c,\dots$ denote three dimensional space indices.
Let $\tilde \eta^{abc}$ be the alternating symbol in space,
$\tilde e = {1\over 3!}\tilde \eta^{abc}e_a^i e_b^j
e_c^k\epsilon_{ijk}$ the determinant of the triad $e^i_a$,
$\tilde E^a_i=\tilde e e^a_i$ the density weighted cotriad and $A^i_a$
the $SU(2)$ connection. The generator of $SU(2)$ rotations is denoted
by $\delta_i$,  so for any $SU(2)$ vector $\omega^j$,
$\delta_i \omega^j=\epsilon^{j}_{\ il}\omega^l$.
The $SU(2)$ covariant derivative is defined by
$D_a=\partial_a+A^i_a \delta_i$. The corresponding curvature is
$F^i_{ab}=\partial_{[a} A^i_{b]}+\epsilon^i_{jk}A^j_a A^k_b$, where
the square brackets denote antisymmetrization without the factor
${1\over 2}$. Let us also introduce, for later purposes
\be
T^i_{ab} := D_{[a}e^i_{b]}.
\ee
%$\partial_{[a}e^i_{b]}+\epsilon^i_{jk} A^j_{[a}e^k_{b]}$.

The constraints of Euclidean general relativity in Ashtekar's
variables are
\begin{eqnarray}
\tilde G_i &\equiv& -D_a \tilde E^a_i=0\label{gauss}\\
\tilde H_a &\equiv& \partial_{[a}A^i_{b]}\tilde
E^b_i-A^i_a\partial_b\tilde
 E^b_i=0\label{diffeo}\\
\tilde{\tilde C} &\equiv& F^i_{ab}\tilde E^a_j \tilde
E^b_k{\epsilon_{i}}^{jk}=0
\label{ham}.
\end{eqnarray}
One often replaces the diffeomorphism constraint by the vector
constraint
\begin{eqnarray}
\tilde V_a\equiv F^i_{ab}\tilde E^b_i=\tilde H_a-A^i_a \tilde G_i.
\end{eqnarray}
which is an intermediate step in our redefinition of the
contraint surface.

Let
\bea
F^i_{\ jk}&=&F^i_{ab}e^a_je^b_k, \ \ \ \ F_i=F^j_{\ ij},\ \ \ \
F=\epsilon_{ijk}F^{ijk}\nn
T^i_{\ jk}&=&T^i_{ab}e^a_je^b_k,\ \ \ \ T_i=T^j_{\ ij},\ \ \ \
T=\epsilon_{ijk}T^{ijk}.
\eea
Consider the covariant derivative
\be D_i=e^a_i D_a.
\ee
Its curvature $F^i_{\ jk}$ and torsion $T^i_{\ jk}$ are given by
\begin{eqnarray}
[D_i,D_j]=F^k_{\ ij}\delta_k-T^k_{\ ij}D_k
\end{eqnarray}

The Bianchi identities following from
$[D_k,[D_i,D_j]]+{\rm cyclic}\ (k,i,j)=0$ are
\begin{eqnarray}
D_k F^j_{\ mn}-F^j_{\ ki}T^i_{\ mn}+{\rm cyclic}\
(k,m,n)&=&0\label{b1}\\
D_k T^j_{\ mn}-T^j_{\ ki}T^i_{\ mn}+ \epsilon^{j}_{\ ki}F^i_{\ mn}+
{\rm cyclic}\ (k,m,n)&=&0. \label{b2}
\end{eqnarray}

The constraint surface defined by the equations
(\ref{gauss})-(\ref{ham}) may equivalently be represented
by the equations
\begin{eqnarray}
T_i=0\label{y1}\\
F_i=0\label{y2}\\
F=0\label{y3}.
\end{eqnarray}
Indeed, the first equation is just the Gauss constraint divided by
$\tilde e$, the second equation is the vector constraint divided by
$\tilde e$ and contracted with $e^a_i$, while the last equation is the
Hamiltonian constraint divided by $(\tilde e)^2$.

Let $\vec\lambda(x),\vec\mu(x)$ be space dependent $SU(2)$ vectors with
$[\vec\lambda,\vec\mu]^i=\epsilon^i_{jk} \lambda^j\mu^k$,
let $\rho(x),\sigma(x)$ be space dependent
scalars and let the smeared version of the
constraint be defined by
\begin{eqnarray}
{\cal T}[\vec\lambda]=\int d^3x\ \tilde e T_i\lambda^i\label{gs2}\\
{\cal F}[\vec\mu]=\int d^3x\ \tilde e F_i\mu^i\label{v1}\\
{\cal C}[\rho]=\int d^3x\ \tilde e F\rho.
\end{eqnarray}
A direct computation using the first of the Bianchi identities
(\ref{b1}) gives for the constraint algebra
\bea
\{{\cal T}[\vec\lambda],{\cal T}[\vec\mu]\}
&=&{\cal T}[{[\vec\lambda,\vec\mu]}]\\
\{{\cal T}[\vec\lambda],{\cal F}[\vec\mu]\}&=&{\cal F}
[{[\vec\lambda,\vec\mu]}]\\
\{ {\cal T}[\vec\lambda],{\cal C}[\rho] \}&=& 0\\
\{{\cal F}[\vec\lambda],{\cal F}[\vec\mu]\}&=&
%\int d^3x\tilde e {-3\over 2}T_jF_i(\lambda^j\mu^i-\lambda^i\mu^j)
%+T_iF^i_{\ kj}\mu^k\lambda^j+F_k T^k_{ji}\lambda^j\mu^i\\
{\cal T}[-\vec F_{jk}\lambda^j\mu^k+{3\over
2}F_i(\lambda^i\vec\mu-\mu^i\vec\lambda)] + {\cal F}[\vec
T_{jk}\lambda^j\mu^k]\nonumber\\
&\equiv&{\cal T}[-\vec F_{jk}\lambda^j\mu^k]+{\cal F}[\vec
T_{jk}\lambda^j\mu^k-{3\over
2}T_i(\lambda^i\vec\mu-\mu^i\vec\lambda)]\\
\{{\cal C}[\rho],{\cal F}[\vec\lambda]\}&=&{\cal T}[{1\over
2}F\rho\vec\lambda+2\rho{{\vec \epsilon}_{i}}^m{F^i}_{km}\lambda^k+
2\rho[\vec\lambda,\vec F]]\nn & & +{\cal F}[-{1\over 2}\rho
T\vec\lambda+2[\vec D\rho,\vec\lambda]+\rho \vec
T_{ij}{\epsilon^{ij}}_k\lambda^k]\\
\{{\cal C}[\rho],{\cal C}[\sigma]\}&=&{\cal C}[-4F^i(\rho
D_i\sigma-\sigma D_i\rho)]\nn
\\ &\equiv&{\cal F}[-4F(\rho\vec D\sigma-\sigma\vec D\rho)]
\eea
Note that the algebra of the modified vector contraints contains
structure
functions, but that these relations contain no derivatives of the
smearing
functions. This is contrary to what happens for the usual representation
(\ref{diffeo}). All the structure functions are $SU(2)$ tensors and
contain no space indices.

\section{Orbit space of $SU(2)$ Yang-Mills theory in the
jet-bundle approach and Wilson loops.\label{2}}
\setcounter{equation}{0}

\subsection{Gauge orbits}

Let us take for simplicity Euclidean space ${\bf R}^3$ as the base space
of the trivial principal bundle $\pi:{\bf R}^3\times
SU(2)\longrightarrow
{\bf R}^3$. An analytic connection $A^i_a$ is a section from
${\bf R}^3$ to $su(2)$ which can be represented by giving all its
partial
derivatives at a point $x_0$. Let us denote by $V^k$ the
space with coordinates
\be
    (A_a^i,\ \p_{b_1}A_a^i,\cdots,\p_{b_k}\cdots\p_{b_1}A_a^i).
 \label{oldc}
\ee
Using a multi-index notation, denote coordinates on $V^k$ collectively
by
$\partial_B A^i_a$, where the order $|B|$ of the multiindex is less than
$k$.
%For example, if
%$B=(0;0)$, $\partial_B A_a^i = A_a^i$, and if
%$B=(1,2,3,3,2,2)$,
%$\p_B={\p^6\over\p x^1(\p x^2)^3(\p x^3)^2}$.
The bundle $\pi:{\bf R}^3\times V^k \longrightarrow {\bf R}^3$ is called
the
$k$-th order jet-bundle and denoted by $J^k$.  A point $\tau$ in $J^k$
has
coordinates
\be
\tau = (x,\ A_a^i,\ \p_{b_1}A_a^i,\cdots,\p_{b_k}\cdots\p_{b_1}A_a^i).
\ee
The spaces $V^k$, and the bundles $J^k$, for $k\in {\bf N}$, form a
projective
family. (For more details see for example \cite{Saunders}.)

A local function $f$ of the connection is by definition a smooth, space
dependent function, which depends only on a finite number of
derivatives of the connection. Hence it belongs to $C^\infty (J^k)$
for some $k$; $f = f(\tau)$.

Gauge transformations of the connection are characterized by giving a
group element $g(x_0)$ at every point $x_0$. If $\tau_i$ are the
Pauli matrices, then $T_j=-{i\over 2}\tau_j$ are  generators of
$SU(2)$, and  gauge transformations act on $A=A^i_aT_idx^a$ as
$A_g=g^{-1}Ag+g^{-1}dg$. If $g$ is of the form $g={\rm
exp}(\epsilon^iT_i)$
with space dependent $\epsilon^i$, the corresponding infinitesimal
gauge transformation are $\delta_\epsilon A^i_a=D_a \epsilon^i$.

The total derivative $d_a$ of a function $f(\tau)$ is
\begin{eqnarray}
d_a f = \p_a f + \p_{Ba} A^i_c {\p f \over\p (\p_B A^i_c)} \ .
\label{p}
\end{eqnarray}
Under infinitesimal gauge transformations
$\delta_\epsilon A_a^i = D_a\epsilon^i$, and $f(\tau)$ changes
according to
\be
\delta_\epsilon f = D_a\epsilon^i {\p f\over \p A_a^i} + \cdots
+ \p_{c_1}\cdots\p_{c_k}(D_a\epsilon^i)
{\p f\over \p(\p_{c_1}\cdots\p_{c_k} A_a^i) }.
\ee
Thus in the jet-bundle $J^k$, infinitesimal gauge transformations are
generated by the family of vector fields
\begin{eqnarray}
\vec X_\epsilon=\partial_B(D_a\epsilon^i)
{\p \over\p (\p_B A^i_a)}\label{vf},
\end{eqnarray}
which are tangent to the fibers $V^k$, and are parametrized by the
functions
$\epsilon^i$. The vector fields on $J^k$ form a module over the algebra
of local functions $C^\infty(J^k)$, and a generating set for
the above family is obtained from (\ref{vf}) by choosing the following
values for the functions $\epsilon^i$ and their derivatives at the point
$x_0$:
\begin{eqnarray}
\epsilon^i= \delta^i_j,\ \partial_a \epsilon^i=0,\cdots,
\partial_{a_1\cdots a_{k+1}}\epsilon^i=0 & \leftrightarrow &
\vec X_j\nonumber\\
\epsilon^i=0,\ \partial_a \epsilon^i=
\delta^b_a\delta^i_j,\cdots,\partial_{a_1\cdots a_{k+1}}
\epsilon^i=0 & \leftrightarrow & \vec X^b_j\nonumber\\
&\vdots & \nonumber\\
\epsilon^i=0,\ \partial_a\epsilon^i=0,\cdots,
\partial_{a_1\cdots a_{k+1}}
\epsilon^i=\delta^{(b_1\cdots b_{k+1})}_{(a_1 \cdots a_{k+1})}
\delta^i_j
& \leftrightarrow & \vec X^{(b_1\cdots b_{k+1})}_j.
\label{cv}
\end{eqnarray}
In other words, an element of the family of vector fields (\ref{vf})
is obtained  by fixing a point
\be
  (x,\ \epsilon^i,\ \p_{a_1}\epsilon^i,\cdots,\p_{a_1}\cdots
  \p_{a_{k+1}}\epsilon^i)
\label{jk+1}
\ee
in the jet-bundle $J^{\prime k+1}$ associated with the sections of the
bundle $\pi:{\bf R}^3\times su(2)\longrightarrow {\bf R}^3$. The above
generating set corresponds to fixing the points defined by the vectors
tangent to each of the coordinate lines in the fibre $V^{\prime k+1}$ of
$J^{\prime k+1}$.

The vector fields
\begin{eqnarray}
\vec X^b_j,\cdots, \vec
X^{(b_1\dots b_{k+1})}_j \label{vf1}
\end{eqnarray}
are in involution. The commutation rules for the entire set $\vec X_j$,
$\vec X^b_j,\cdots$, $\vec X^{(b_1\cdots b_{k+1})}_j$ are summarized in
the
relation
\begin{eqnarray}
[\vec X_\epsilon,\vec X_\eta]=\partial_{B}[D_a(\epsilon^i_{\
jk}\epsilon^j\eta^k)]
{\p \over\p (\p_{B}A^i_a)}. \label{cr}
\end{eqnarray}
The involution property is deduced from this by
choosing the canonical values (\ref{cv}) for $\epsilon^j$, $\eta^k$ and
their
derivatives.

By Frobenius theorem, the set of vector fields $\vec X_j$,
$\vec X^b_j,\dots$, $\vec X^{(b_1\dots b_{k+1})}_j$ is integrable, and
hence
tangent to finite dimensional integral submanifolds of the fibers $V^k$.
These submanifolds are just the gauge orbits ${\cal G}^k$. The
collection
of maximal dimensional gauge orbits defines a foliation of the fibers
$V^k$;
the gauge orbits are the leaves of this foliation.

\subsection{Orbit space}

Let us now investigate the linear independence of the vector fields
(\ref{vf1}) in order to study the structure of the space of orbits
$V^k/{\cal G}^k$. Consider the following functions on $V^k$ :
\begin{eqnarray}
& & A_a^i,\ \p_{(b_1}A^i_{a)},\cdots
,\ \p_{(b_k}\cdots \p_{b_1} A^i_{a)},\label{y5}\\
& & F_{b_1a},\ D_{(b_2}F_{b_1)a},\cdots
,\ D_{(b_k} \cdots D_{b_2} F^i_{b_1)a},\label{x5}
\end{eqnarray}
These functions can be taken as new coordinates on $V^k$
\cite{DVHTV,Tor}.
In the Abelian case, this change of coordinates corresponds exactly to
separating  the old coordinates  $\p_{b_k}\cdots\p_{b_1}A_a$
(\ref{oldc})
into pieces symmetrized and anti-symmetrized on $a$ and $b_i$,
for any $1\le i\le k$. In the non-Abelian case the basic idea is the
same,
although the details are different due to the commutator in $F_{ab}^i$.

In the new coordinates, the family of vector fields (\ref{vf}) is
\bea
\vec X_\epsilon &=& \sum_{l=0}^k\ [\partial_{(b_l\cdots b_1}D_{a)}
\epsilon^i {\p\over\p (\p_{(b_l\cdots b_1}A^i_{a)})}
\nonumber\\
& &+\epsilon^i_{\ jk}D_{(b_{l}}\cdots D_{b_2} F^j_{b_1) a}
\epsilon^k{\p \over\p (D_{(b_{l}}\cdots D_{b_2} F^i_{b_1)a})}].
\eea
It is then straightforward to check that an equivalent generating set
to (\ref{cv}) is obtained by making the following choice for the
gauge parameters $\epsilon_i$ and their derivatives :
\begin{eqnarray}
\epsilon^i= \delta^i_j,\ D_a
\epsilon^i=0,\ \dots,\
\partial_{(a_1\dots a_{k}}D_{a_{k+1})}\epsilon^i=0 &
\leftrightarrow &
\vec Y_j\nonumber\cr
\epsilon^i=0,\ D_a \epsilon^i=
\delta^b_a\delta^i_j,\ \dots,\ \partial_{(a_1\dots a_{k}}D_{a_{k+1})}
\epsilon^i=0 & \leftrightarrow &
{\partial\over\partial A^j_b}\nonumber\cr
&\vdots &\nonumber\cr \epsilon^i=0,\ D_a \epsilon^i=0,\ \dots,\
\partial_{(a_1\dots a_{k}}D_{a_{k+1})}
\epsilon^i=\delta^{(b_1\dots b_{k+1})}_{(a_1
\dots a_{k+1})} \delta^i_j & \leftrightarrow &
{\partial\over\partial(\partial_{(b_1\dots b_{k}}
A^j_{b_{k+1})})},
\end{eqnarray}
where
\begin{eqnarray}
\vec Y_k=\sum_{l=0}^k\ [\epsilon^i_{\ jk}D_{(b_{l}}\cdots
D_{b_2}F^j_{b_1) a}{\p \over\p (D_{(b_{l}}\cdots D_{b_2}F^i_{b_1)a})}].
\end{eqnarray}

This new choice of values for the gauge parameters corresponds to the
following situation. We have the Whitney sum bundle $J^k\oplus
J^{\prime k+1}$, whose fiber consists of $V^k\oplus V^{\prime
k+1}$. In this direct sum, the new choice of gauge parameters
corresponds
to a change of coordinates in the second factor, from the old
coordinates
(\ref{jk+1}) to the new ones
\be
(x,\ \epsilon^i,\ D_{a_1}\epsilon^i,\cdots,\ \p_{(a_1}\cdots
\p_{a_k}D_{a_{k+1})}\epsilon^i ).
\ee
The associated generating set for the gauge orbits
is obtained by fixing, in the second factor of the sum, those points
which
are determined by the vectors tangent to the new coordinate lines.

The question of linear independence is now reduced to the
investigation of the linear independence of the three vector fields
$\vec
Y_j$, since the other vector fields, being tangent to different
coordinate
lines, are obviously independent. Alternatively, one sees that the
coordinates (\ref{y5}) are coordinates purely along the gauge orbits,
while
the remaining coordinates transform under the adjoint action of the
group.
This is reminiscent of what happens if one considers holonomies around
closed loops as the basic variables of the theory, which also transform
under the adjoint action. This analogy will be made more precise in the
last part of this section.

As an example consider the space $V^1$. The
coordinates on $V^1$ are
\be
(A_a^i,\ \p_{(b_1}A^i_{a)},\ F^i_{{b_1}a} ).
\ee
In two spacetime dimensions the three vector fields $\vec Y_i$ are
\be
\vec Y_k=\epsilon^i_{\ jk}F^j_{01}
{\partial\over\partial F^i_{01}}.
\ee
These are the field space analogs of the usual angular momentum
generators
(for one particle) in three dimensions $\epsilon_{ij}^{\ \ k}x^j\p_k$.
On the origin $\vec F_{01}=0$, and all three vector fields vanish,
while for $\vec F_{01}\ne 0$, there is one relation between them.
Their orbits are the $2$-dimensional spheres centered at the origin
in ${\bf R}^3$ with coordinates $F^i_{01}$.

In more than two spacetime dimensions, or for $V^k$ with $k>1$, the
three vector fields $\vec Y_k$ are of the form
\be
\vec Y_k=\epsilon^i_{\ jk}x^j_S
{\p \over\p x^i_S},
\ee
where we have used $x^i_S$ to denote the coordinates
$D_{(b_{l}}\cdots D_{b_2}F^i_{b_1)a}$. The range $N$ of the index
$S=1,2,\cdots,N$ depends on the spacetime dimension and $k$. Thus
$\vec Y_k$ looks like the sum of the angular momentum generators
of $N$ particles: $\vec Y_k = \vec Y_k^{(1)}+\cdots+ \vec Y_k^{(N)}$.
In the generic situation, all of the $N$ particles will not lie on a
line through the origin, and therefore the orbits of the three
$Y_k$ will be three-dimensional.

\subsection{Wilson loops}

Any gauge invariant polynomial or formal power series on $J^k$ can be
written
as a power series in the $x^i_S$, where all the internal indices are
tied up
with the invariant
tensors $\delta_{ij}$ and $\epsilon_{ijk}$. This follows from an
analysis of the BRST cohomology (see for instance \cite{DVHTV}).
On the
other hand, it is well
known that Wilson loops are non-local gauge invariant objects, and that
their knowledge, for all loops, fixes the gauge potentials up
to a gauge transformation \cite{Giles}. The object of the following is
to
show that analytic Wilson loops can be written as a formal power series
of invariant monomials in the coordinates $x^i_S$.

First of all, it is straightforward to see
that holonomies can be described as a formal power series on $J^\infty$.
Consider a path $\gamma$ in $\bf R^3$, with base point
$x_0$. Divide $\gamma$  into $n+1$ segments given by displacement
vectors
$\Delta x^a_k$, $0\le k\le n$. Then the discretized holonomy is
\bea
H^D_\gamma[A] &=& [1-A_a^i(x_0)\tau^i\Delta x_0^a]\
[1-A_a^i(x_1)\tau^i\Delta x_1^a] \cdots [1-A_a^i(x_n)\tau^i\Delta x_n^a]
\nn
&=& \prod_{k=0}^n\ [1-A_a^i(x_k)\tau^i \Delta x_k^a],
\label{disw}
\eea
with the continuum limit given by
\be
H_\gamma[A] = \lim_{\Delta x^a\to 0 \atop n\to \infty} H^D_\gamma[A].
\ee
To rewrite $H^D_\gamma[A]$ as a polynomial on $J^n$, we have to
express each $A_a^i(x_k)$ as a function of derivatives of $A_a^i$
evaluated at the base point $x_0$. The answer is simply
\bea
A_a^i(x_1) &=& A_a^i(x_0) + (\p_{b_1}A_a^i)(x_0)\Delta x_0^{b_1}
\nn
A_a^i(x_2) &=& A_a^i(x_0)
+ (\p_{b_1}A_a^i)(x_0)\ (\Delta x_0^{b_1}+ \Delta x_1^{b_1})
+ (\p_{b_2}\p_{b_1}A_a^i)(x_0)\Delta x_0^{b_1}\Delta x_1^{b_2}
\nn
{\rm etc.} &&
\eea
Each term in (\ref{disw}) may now be rewritten in the new coordinates
(\ref{y5})-(\ref{x5}). In the continuum limit, we get a formal power
series
on $J^\infty$.

We now want to show that, for closed loops,
the coordinates (\ref{y5}) do not appear and that, if one takes
the trace to obtain a gauge invariant functional, the Wilson loop,
the remaining coordinates (\ref{x5}) are contracted on their
internal indices with invariant tensors.
This can be deduced as follows. The gauge invariance of the Wilson loop
implies the gauge invariance of its discretized version, which can be
described, as seen above, as a polynomial on $J^n$. Since
this polynomial is gauge invariant,
it must be an invariant polynomial in the $x^i_S$ \cite{DVHTV}.
Alternatively, we can give the
following constructive proof.

Let us adopt the conventions of the non abelian Stokes theorem
\cite{Ar}, i.e., take a surface $\Sigma$ in ${\bf R}^3$ defined by
analytic functions $x^a=f^a(s,t)$ with $0\leq
s,t\leq 1$ and $f^{\prime a}=\p f^a/ \p s$, $\dot
f^a=\p f^a/\p t$. Let $h(s,t)$ be the holonomy along
the curve $f^a(s^\prime,t)$, $0\leq s^\prime\leq s$ at fixed $t$ and
$g(s,t)$ the holonomy along the curve $f^a(s,t^\prime)$, $0\leq
t^\prime\leq t$ at fixed $s$. Note that in our conventions, the
holonomy around a path $\gamma$ is defined by
$H_\gamma={\rm Pexp}(\int_\gamma -A^i_aT_idx^a)$.
Following \cite{Ar}, we divide the square $[0,1]\times[0,1]$ into $nm$
rectangles with sides ${1\over n}, {1\over m}$. The
holonomy around the boundary $\partial\Sigma$ is given by
\be
H(\partial \Sigma)=\lim_{n,m\rightarrow\infty}H_{n,m}(\partial
\Sigma)\label{f}
\ee
with
\be
H_{n,m}(\partial\Sigma)={\cal P}_{s,t}\prod_{l,k=0}^{n-1,m-1}Sp(l,k)
\label{f1}.
\ee
In this equation ${\cal P}_{s,t}$ denotes the ordering which puts a
matrix with the large value of the first argument to the right and,
for identical first arguments it puts the one with the smaller second
argument to the right, while $Sp(l,k)$ is the holonomy around
the spoon loop
with bowl based at $f^a(l,k)$ :
\bea
Sp(l,k) &\equiv& h^{-1}({l\over n},0)\ g^{-1}({l\over n}, {k\over m})\
[I+{1\over nm}T_i F^i_{ab}f^{\prime a}\dot f^b(f({l\over n},{k\over m}))
\nn
&& + o({1\over nm})]\ g({l\over n},{k\over m})\ h({l\over n},0).
\eea
Following the reasoning in \cite{Loos}, we find that this holonomy
reduces to
\bea
Sp(l,k)
&=& h^{-1}({l\over n},0)\ g^{-1}({l\over n},{k-1\over m})\
[I+{1\over nm}T_i f^{\prime a}\dot f^b(f({l\over n},{k\over m}))
\nn
& & \{1+{1\over m}\dot f^c D_c\}\
F^i_{ab}(f({l\over n},{k-1\over m}))
\nn
& & + o({1\over nm})]\ g({l\over n},{k-1\over m})\ h({l\over n},0)
\nn
&=& h^{-1}({l\over n},0)\
[I+{1\over nm}T_i f^{\prime a}\dot f^b(f({l\over n},{k\over m}))
 \{ 1+{1\over m}\dot f^{c_k} D_{c_k}\}
\nn
& & \cdots\{1+{1\over m}\dot f^{c_1} D_{c_1}
\}F^i_{ab}(f({l\over n},0))+o({1\over nm})]\ h({l\over n},0)
\nn
&=& [I+{1\over nm}T_i f^{\prime a}\dot f^b(f({l\over n},{k\over m}))
\{ 1+{1\over n}f^{\prime b_l} D_{b_l}\}\cdots
\{1+{1\over n} f^{\prime b_1} D_{b_1}\}
\nn
& &\{ 1+{1\over m}\dot f^{c_k} D_{c_k}\}\cdots\{1+{1\over m}
\dot f^{c_1} D_{c_1}\} F^i_{ab} (f(0,0))+o({1\over nm})]
\nn
&=& [I+{1\over nm}T_i f^{\prime a}\dot f^b(f({l\over n},{k\over m}))
 \{{\rm exp}({l\over n}  f^{\prime c} D_c)\}
 \nn
& & \{{\rm exp}({k\over m}\dot f^c D_c)\}F^i_{ab}(f(0,0))
+o({1\over nm})].
\eea
Injecting this result into formula (\ref{f1}), and using the fact that
${\rm Tr} (T_{i_1}\dots T_{i_k})$ is an invariant tensor under the
adjoint action of $su(2)$, (it is a linear
combination of $\delta_{ij},\epsilon_{ijk}$ and their contractions),
we find indeed that the Wilson loop ${\rm Tr} H(\partial\Sigma)$ can be
written as a power series depending on the field strengths and all their
symmetrized covariant derivatives $D_{(b_k}\cdots D_{b_{2}}F^i_{b_1)a}$,
$k=1,\cdots,\infty$ evaluated at the base point of the loop.
There are contractions on the group indices with
invariant $su(2)$ tensors, and on
the spatial indices with coefficients characterizing the loop
$\partial\Sigma$. It also follows from this derivation that, at every
finite level of approximation, the Wilson loop can be described as a
local
function, depending on invariant monomials in the $x^i_S$,
and it is only when one takes the continuum limit
$n, m\rightarrow \infty$ that
it becomes a function involving an infinite number of derivatives.

\section{Construction of the geometrical representation of the
constraint
surface}
\setcounter{equation}{0}

So far we have considered the jet space associated with $SU(2)$
Yang-Mills theory and considered an alternative set of
coordinates on this space. This set of coordinates was defined in
such a way as to
isolate pure gauge directions. In this section we describe a
similar change of coordinates on the jet space of
Euclidean canonical general relativity in Ashtekar's variables. This
is again designed to isolate pure gauge directions, for
the gauge orbits generated by the kinematical constraints.
This
leads to the geometrical representation of the constraint surface.
The general strategy and theorems on
how to do this are explained in Ref. \cite{B}, and have already been
used in the context of Lorentzian tetrad gravity in Ref. \cite{BBH3}.
Similar ideas for gravity in Ashtekar's variables in the Lagrangian
approach have been discussed in Ref. \cite{Mor}.

The field content of the theory is given by the $SU(2)$ connection
$A^i_a$ and the dreibein $e^i_a$. In addition, there are the gauge
parameters
for the Gauss and diffeomorphism constraints $\eta^i$ and $\eta^a$.
We must consider the jet-space of all these fields.  Since we will not
consider the gauge orbits generated by the Hamiltonian constraint, we
do not concern ourselves with the associated gauge parameter.

The smeared constraints
\begin{eqnarray}
\int d^3x\ (\tilde G_i\eta^i+\tilde H_a\eta^a)\label{sm}
\end{eqnarray}
generate the gauge transformations
\begin{eqnarray}
\gamma A^i_a &=& D_a\eta^i+ L_\eta A^i_a \nn
\gamma e^i_a &=& -\eta^k\delta_ke^i_a + L_\eta e^i_a\label{gtr}
\end{eqnarray}
where the Lie derivative $L_\eta$ is given by $L_\eta
A^i_a=\eta^c\p_c A^i_a + A^i_c \p_a\eta^c $, and similarily
for $e^i_a$. The gauge parameters $\eta^i,\eta^a$
are taken to be commuting in this section, but in the BRST
context, they are replaced by anticommuting ``ghosts''.

Following Sec. 3 of \cite{BBH3}, we consider the set of coordinates
\bea
&&\p_{(a_l}\cdots\partial_{a_1}A^i_{b)},\ \ \
\p_{(a_l}\cdots\p_{a_1}e^i_{b)}\label{x1}\\
&&D_{(i_{l}}\cdots D_{i_2}F^k_{\ i_1)j},\ \ \
D_{(i_{l}}\cdots D_{i_2}T^k_{\ i_1)j}\label{x2}\\
&&\hat C^i=\eta^i+\eta^a A^i_a,\ \ \
\hat\xi^i=e^i_a\eta^a,\label{x3}\\
&&\p_{(a_l}\cdots\p_{a_2}K^i_{a_1)},\ \ \
\p_{(a_l}\cdots\p_{a_2}L^i_{a_1)},\label{x4}
\eea
where $l=0,\dots,k$. The  $K^i_a$ and
$L^i_a$ are gauge parameters replacing the derivatives of $\eta^i$
and $\eta^a$,
and are defined by the combinations appearing on the r.h.s of the gauge
transformations (\ref{gtr}): $K_a^i\equiv \gamma e_a^i$, $L_a^i\equiv
\gamma A_a^i$.

By following the same reasoning as in the previous section, that is,
rewriting the vector fields generating the gauge transformations in
the new coordinate system, and then showing that a generating set is
obtained by giving canonical values to the combinations of
gauge parameters  (\ref{x3}) and (\ref{x4}),
we can see that the coordinates (\ref{x1}) are purely
along the gauge orbits. Indeed, giving canonical values to the
parameters
(\ref{x4}), one finds that the generating set contains the vector fields
tangent to these coordinate lines.

An alternative way to see this is the following: If, for example,
$f=f(\p_{(a} e^i_{b)}, \p_{(a} A^i_{b)})$, then
\be
f + \gamma f = f +  {\p f \over \p_{(a}e^i_{b)}} \p_{(a}K_{b)}^i
+ {\p f\over \p_{(a}A^i_{b)} } \p_{(a}L_{b)}^i =
f( \p_{(a}e^i_{b)}+ \p_{(a}K_{b)}^i, \p_{(a}A^i_{b)} + \p_{(a}L_{b)}^i),
\ee
for infinitesimal transformations,
which shows that the gauge transformations are just translations by
(\ref{x4}) along the coordinate lines of  $\p_{(a} A^i_{b)}$ and
$\p_{(a} e^i_{b)}$.

The coordinates (\ref{x2}) are therefore the only ones that are partly
transversal to the gauge orbits. Denoting these collectively by ${\cal
T}^r$,
we see that they transform among themselves
with the parameters (\ref{x3}) alone according to
\begin{eqnarray}
\gamma {\cal T}^r=-\hat C^k\delta_k {\cal T}^r + \hat \xi^k D_k
{\cal T}^r.
\end{eqnarray}
If one now expresses the sum of the smeared constraints (\ref{sm})
in the new coordinate systems, one finds the expression
\begin{eqnarray}
\int d^3x \tilde e (F_i \hat \xi^i-T_i\hat C^i).
\end{eqnarray}
This implies the geometric representation (\ref{y1})-(\ref{y2})
of the constraint surface in terms of the ${\cal T}^r$ alone,
in agreement with the general theorem of \cite{B}.

It is well known that first class constraints play a double role,
the first as generators of gauge transformations,
the second as the restrictions which give
physically acceptable initial data. Having considered the first aspect,
we now turn to the constraints (\ref{y1})-(\ref{y3}) as restrictions.

%It is well known that first class constraints play a double role, the
%first as generators of gauge transformations, and the second as the
%restrictions
%which give physically acceptable initial data. Having considered the
%first
%aspect,
%we now turn  to the constraints (\ref{y1})-(\ref{y3}) as restrictions.
%We show
%in the following that these can be written in terms of the coordinates
%${\cal T}^r$, a fact which is in agreement with the general theorem of
%\cite{B}.

To do this explicitly, it is neccessary to further split
the coordinates ${\cal T}^r$. We decompose the dual
$\epsilon^{ijl}F^k_{\ ij}$
of $F^k_{ij}$ into a trace free symmetric part, a trace, and an
antisymmetric part,
\be
F^k_{\ ij} = \epsilon_{ijl} F_T^{(kl)}
+ {1\over 6}\epsilon_{ij}^{\ \ k} F
+ {1\over 2} \epsilon_{ijl}\epsilon^{klm} F_m,
\label{decom}
\ee
where
$F_T^{(kl)}={1\over 2}\epsilon^{ij(k}F_{ij}^{l)}-{1\over
6}\delta^{kl}F$.
{}From this decomposition it is clear that the only non-gauge and
non-constraint coordinates on the jet space are the first and second
terms in (\ref{decom}), their corresponding symmetrized derivatives,
together with analogous coordinates from the
identical decomposition of $T^k_{\ ij}$. Note also that in this
decomposition,
the third term is just the diffeomorphism constraint. As we will see in
the next
section, these remaining coordinates turn out to be useful in
classifying
spatial-diffeomorphism and Gauss invariant observables.

\section{Classification of local conservation laws \label{la}}
\setcounter{equation}{0}

Consider the four-dimensional generally covariant $SU(2)$ gauge field
theory with action
\be
S= \int {\rm Tr} (e \wedge e \wedge F).
\label{hkact}
\ee
This action is identical in form to that for general relativity except
for
the gauge group, which is $SU(2)$ instead of $SL(2,C)$.  The Hamiltonian
description has an identically vanishing first class Hamiltonian
\cite{HK},
and two first class constraints, which are the Gauss and the
diffeomorphism
constraints (\ref{gauss})-(\ref{diffeo}), or equivalently, their
covariant
versions (\ref{y1})-(\ref{y2}). The geometrical coordinates presented in
the
last section are therefore very useful in discussing the local
conservation laws of this model.

Local conserved currents $\tilde{j}^a$ are vector densities constructed
from local functions of the fields and their derivatives which satisfy
$d_a \tilde{j}^a = 0$ when the equations of motion hold, and where
$d_a$ is defined as in (\ref{p}) above, but includes all the fields in
the theory. The dual description is in terms of horizontal forms, which
are
defined to be forms on spacetime, or on space, with coefficients
that are local functions. On spacetime, local functions also involve
time derivatives, whereas on space, they involve only spatial
derivatives.
The horizontal exterior derivative is defined by $d\equiv dx^a d_a$,
where
the index $a$ goes from 0 to 3 for spacetime, or from 1 to 3 for
space. Thus, in $n$ spacetime dimensions, we define the $(n-1)$-form
$j_{ a_1\cdots a_{n-1} }:= \epsilon_{a_1\cdots a_n} \tilde{j}^{a_n}$,
and the conservation equation becomes $dj=0$, when the equations of
motion are satisfied.

Let $\Sigma$ denote the surface defined by the equations of motion and
their derivatives, and $\tilde \Sigma$ the surface defined by the
constraints
and their spatial derivatives. Then, for theories in $n$ spacetime
dimensions,
the vector space of local conservation laws is the  equivalence classes
of
horizontal ($n-1$)-forms $j$ on spacetime which
satisfy $dj=  0$ on $\Sigma$, and where two such forms are considered
equivalent if they differ on $\Sigma$ by the exterior horizontal
derivative of a horizontal $n-2$ form $k$: $j\sim j+dk$ on $\Sigma$.
In what follows, and in the
Appendix, we consider only ``dynamical'' conservation laws and
cohomology groups, and not ``topological'' ones.
The latter come from non-triviality of the triad manifold ($\tilde
e\neq
0$).
Following \cite{BBH3}, one can easily generalize the subsequent
considerations to include these additional conservation laws and
cohomology groups.

One can prove (see Appendix) that the vector space of local
conservation
laws of a diffeomorphism invariant gauge field theory
with vanishing Hamiltonian is isomorphic to the direct sum of the
following
two vector spaces: (i) the vector space of conservation laws in space
associated with $\tilde{\Sigma}$, and (ii) the vector space of
equivalence classes of horizontal $(n-1)$-forms in space which are
invariant on $\tilde \Sigma$ under the transformations generated by the
constraints, up to exact $(n-1)$-forms in space; the equivalence
relation sets two such forms to be equal if they differ on
$\tilde{\Sigma}$ by the horizontal exterior derivative $d$ of a
horizontal ($n-2$)-form in space.
Let us call this last space ${\cal O}$.

In the present case, one can prove (Appendix) that
the former space is trivial. To
describe ${\cal O}$, we use the decomposition (\ref{decom}): The
non-gauge
coordinates (\ref{x2}) decompose into sets
\begin{eqnarray}
&& D_{(i_l}\cdots D_{i_1}\epsilon_{\ i)j}^kF,\ D_{(i_l}\cdots
D_{i_1}\epsilon_{\ i)j}^kT\\
&& D_{(i_l}\dots D_{i_1}\epsilon_{i)jl}\ F^{(kl)}_T,
\ D_{(i_l}\dots D_{i_1}\epsilon_{i)jl}\ T^{(kl)}_T,\\
&& D_{(i_l}\dots D_{i_1}\epsilon_{i)jl}\epsilon^{lkm}F_m,
\ D_{(i_l}\dots D_{i_1}\epsilon_{i)jl}\epsilon^{lkm}T_m
\label{4.3}.
\end{eqnarray}
The first two sets, denoted collectively by ${\cal T}^{\prime r}$, do
not vanish on the constraint surface, while the last obviously does.
Consider functions
$f({\cal T}')$ satisfying $\delta_i f({\cal T}')=0$.  Denote by
$L({\cal T}')$ the equivalence classes of all such functions under the
equivalence relation $f \sim f + D_iM^i$, where $M^i({\cal T}')$
transforms
like a vector under $SU(2)$ transformations.

With these notations, it follows from the Appendix that
${\cal O}$ is described by linear combinations of the
Chern-Simons functional
\begin{eqnarray}
\int \  Tr(AdA+{2\over 3}A^3)
\end{eqnarray}
and
the functionals
\begin{eqnarray}
\int d^3x\ \tilde e L({\cal T}^\prime).
\end{eqnarray}

\section{Models from the Hamiltonian constraint function}
\setcounter{equation}{0}

Consider the Hamiltonian constraint function in the three-dimensional
action
\begin{eqnarray}
S[A_a^i,e^b_j]=\int d^3x\ \tilde eF.\label{top}
\end{eqnarray}
The corresponding field equations are
\begin{eqnarray}
{\delta S\over \delta A_a^i(x)}\equiv\tilde e\epsilon_i^{\
jk}(-T_ke^a_j-T^l_{jk}e^a_l)(x)=0\\
{\delta S\over \delta e^a_i(x)}\equiv\tilde e(-e^i_a F+2\epsilon_l^{\
ik}F^l_{ak})(x)=0.
\end{eqnarray}
Contracting both equations with $e^i_a$ yields $F=0=T$. Inserting
the definitions of $F^i_{\ jk}$ and $T^i_{\ jk}$ in terms of their duals
gives the final result $F^i_{jk}=0=T^i_{jk}$. This means that the
field equations following from  (\ref{top}) require all the local gauge
invariant quantities that can be built out of the connection $A_a^i$
and the triad $e^i_a$ to vanish. It is in this sense that the action
(\ref{top}) plays the same role for the theory based on the $A^i_a$
and $e^i_a$ with the covariant derivative $D_i$ as the Chern-Simons
action plays for the theory based on $A^i_a$ alone with the covariant
derivative $D_a$. The theory given by (\ref{top}) is in fact
three-dimensional (Euclidean) Einstein gravity, better known
in the form
\begin{eqnarray}
S[A_a^i,e_b^j]=\int  {\rm Tr} (e \wedge F).
\end{eqnarray}

One can also write down a four-dimensional action involving a function
similar to the Hamiltonian constraint. Consider the following action
made from an $SU(2)$ connection $A_\mu^i$, dreibein $e^{\mu i}$, and a
scalar density $\tilde{\Phi}$ :
\be
S= \int d^4x\ \tilde{\Phi} e^{\mu i} e^{\nu j}F_{\mu\nu}^k
\epsilon_{ijk}.
\ee
The spacetime metric $g^{\mu\nu}=e^\mu_ie^{\nu i}$ is degenerate, with
degeneracy direction given by the 1-form
$V_\mu = \tilde{\Phi}\epsilon_{\mu\nu\alpha\beta}\epsilon_{ijk}e^{\nu
i}
e^{\alpha j}e^{\beta k}$; $V_\mu g^{\mu\nu}=0$. The field equations
are
\bea
e^{\mu i} e^{\nu j}F_{\mu\nu}^k \epsilon_{ijk} &=& 0, \nn
\epsilon_{ijk} D_\mu (\tilde{\Phi} e^{\mu i} e^{\nu j}) &=& 0, \nn
\tilde{\Phi}  e^{\nu j}F_{\mu\nu}^k \epsilon_{ijk} &=& 0.
\eea
It is clear that the first equation, which is like a Hamiltonian
constraint, is identically satisfied  as a consequence of the third
equation. We suppose that $\tilde \Phi$ is different from zero
everywhere.
 The dynamics is therefore determined entirely by the
latter two equations. Both these equations have spatial projections
which are the constraints of the theory. The standard 3+1 decompostion
of the action reveals that the constraints are in fact just the
Gauss and spatial-diffeomorphism constraints
(\ref{gauss})-(\ref{diffeo}).
Indeed,
the 3+1 form of the action is
\be
S = \int dt \int d^3x\  [ \Pi^a_i \dot{A}_a^i + A_0^i D_a \Pi^{ai}
                  + \tilde{\Phi} e^{ai}e^{bj} F_{ab}^k \epsilon_{ijk} ]
\ee
where  $\Pi^a_k = 2\tilde{\Phi} e^{0i}e^{aj}\epsilon_{ijk}$. We can
rewrite
$\tilde{\Phi} e^{ai}e^{bj}\epsilon_{ijk}$ entirely in terms of $\Pi^a_k$
and
a Lagrange multiplier as follows:
\bea
{1\over 2\tilde{\Phi}} (e^{bl} e^0_l) \Pi^a_k  &=& e^{bl} e^0_l e^0_i
e^a_j
\epsilon^{ijk}
\nn
&=& e^{bl} ( e^T_{il} + { \delta_{il}\over 3}\ e^{0m}e^0_m) e^a_j
\epsilon^{ijk}
\nn
&=& e^{bl} e^T_{il} e^a_j \epsilon^{ijk} + {1\over 3 } e^{0m}e^0_m
e^b_i e^a_j
\epsilon^{ijk},
\eea
where $e^T_{il}$ is the symmetric trace free part of $e^0_l e^0_i$. So
finally
\be
e^a_i e^b_j \epsilon^{ijk} = -{3 e^{bl} e^0_l \over
2\tilde{\Phi}e^{0m}e^0_m}
\Pi^{ak}
                            + {3 \over e^{0m}e^0_m} e^{bl} e^T_{il}
e^a_j
\epsilon^{ijk}.
\ee
Substituting this into the 3+1 action, the second piece contracted with
$F_{ab}^k$ vanishes, and we get
\be
S = \int dt \int d^3x\  [ \Pi^a_i \dot{A}_a^i + A_0^i D_a \Pi^{ai}
                  - N^b \Pi^a_k F_{ab}^k ],
\ee
with the shift function $N^a$ defined by
\be
N^a = {3 e^{al} e^0_l \over 2 e^{0m}e^0_m }.
\ee
The Hamiltonian constraint vanishes identically, a fact which is already
clear from the first field equation. This theory is therefore locally
equivalent to (\ref{hkact}).

\section{Conclusion}

At the price of not using the canonical momenta given by the density
weighted cotriad alone, but working instead with both the triads and the
cotriads, we have shown that there is a natural covariant derivative
acting on $su(2)$ tensors in canonical
Euclidean general relativity in Ashtekar's variables. The appealing
feature of the associated tensor calculus is that the constraints
become algebraic restrictions on
the torsion and the curvature of this covariant derivative. This gives
the canonical formulation a geometrical flavor analogous to the one of
the original
Lagrangian Einstein equations.

Furthermore, all $SU(2)$ and diffeomorphism invariant integrated local
quantities can be classified. Their integrands are shown to be given
either by the Chern-Simons Lagrangian or by the
invariant volume element times $SU(2)$
invariant functions of covariant derivatives of the curvature and the
torsion. This classification is achieved through the computation of the
BRST cohomology of the Husain-Kucha\v r model.

These results help to address the following questions of \cite{HK}~:

(i)~On the basis of our classification of local observables, one can
look
for a complete set of observables which are in involution to decide on
the one hand if the  model is integrable and, on
the other hand to try to quantize the model  in a more traditional way,
to be compared with the loop quantization of \cite{ALMMT}~;

(ii)~A
complete computation of the local BRST cohomology including the
Hamiltonian
constraint would clearly show the difference the inclusion of this
constraint makes on the level of local integrated observables. In fact
it is not really necessary to do the computation, since we can use the
results of \cite{Torre}, which state (in the context of metric
gravity) that there are no local gauge invariant
observables. Consequently, the inclusion of
the Hamiltonian constraint as a generator for gauge symmetries is
responsible for removing all local observables.

\section*{Acknowledgements}
G. B. would like to thank Friedemann Brandt for
useful discussions and the Fonds National Belge de la Recherche
Scientifique for financial support. The work of V. H. was partly
supported by NSF grant PHY 93-96246, the Eberly Research Funds of the
Pennsylvania State University, and the Natural Science and
Engineering Research Council of Canada.

\section*{Appendix: Local BRST cohomology}
\renewcommand{\theequation}{A.\arabic{equation}}
\setcounter{equation}{0}

In this Appendix we give the computation of the local BRST
cohomology associated with the theory given by the action (\ref{hkact}).
The analysis follows closely that of \cite{BBH3}, where the local
BRST cohomology of the Einstein-Yang-Mills theory is analysed.

Let us introduce besides the diffeomorphism and the $SU(2)$ ghosts
$\eta^a$ and $\eta^i$ of (\ref{gtr}), their canonically conjugate
ghost momenta ghost momenta ${\cal P}_a$ and ${\cal P}_i$. The BRST
charge \cite{BFV} of the model (\ref{hkact}) is given by
\begin{eqnarray}
\Omega=\int d^3x\ ( \tilde G_i\eta^i+\tilde H_a\eta^a-{1\over 2} {\cal
P}_k
\epsilon^k_{\ ij}\eta^i\eta^j+{\cal P}_i\eta^a\partial_a\eta^i
+{\cal P}_b\eta^a\partial_a\eta^b).
\end{eqnarray}
The nilpotent BRST transformations $s_\omega$ of the fields are
generated by taking the Poisson
bracket of the fields with $\Omega$. As in \cite{BBH3}, it can then be
verified that a new coordinate system for the jet-bundles associated
to the fields $A^i_a, e^i_a$, the ghosts $\eta^i,\eta^a$ and the ghost
momenta ${\cal P}_i,{\cal P}_a$ is given by the coordinates (\ref{x1})
collectively denoted by $U^s$ and their BRST variations $V^t=s_\omega
U^t$, the ${\cal T}^r$, the $\hat C^i, \hat \xi^i$ and the
${\cal T}^*_s\equiv D_{(i_l}\cdots D_{i_1)}\hat C^*_i,
D_{(i_l}\cdots D_{i_1)}\hat
\xi^*_i$, $l=0,1,\cdots$ with $\hat C^*_i={1\over \tilde e}{\cal
P}_i$
and $\hat \xi^*_i={1\over \tilde e}e^a_i({\cal P}_a-A^k_a{\cal P}_k)$.

The BRST transformations in the new coordinate system act by
convention from the right and are given by
\bea
 s_\omega U^t&=&V^t,\ s_\omega V^t=0
\\
s_\omega{\cal T}^r&=&-\delta_k {\cal T}^r\hat C^k+D_k
{\cal T}^r\hat \xi^k
\\
 s_\omega \hat C^i&=&{1\over 2}\epsilon^i_{jk}\hat C^j\hat C^k-\hat
F^i,
s_\omega \hat \xi^i=-\delta_k\hat \xi^i\hat C^k-\hat T^i
\\
 s_\omega D_{(i_l}\dots D_{i_1)}\hat C^*_i&=&D_{(i_l}\dots D_{i_1)}T_i
 -\delta_k D_{(i_l}\dots D_{i_1)}\hat C^*_i\hat C^k
 \nn
&& + D_k D_{(i_l}\cdots
D_{i_1)}\hat C^*_i \hat \xi^k,\label{b5}
\\
s_\omega D_{(i_l}\cdots D_{i_1)}\hat \xi^*_i&=&-D_{(i_l}\dots
D_{i_1)}F_i
 -\delta_k D_{(i_l}\cdots D_{i_1)}\hat \xi^*_i\hat C^k
\nn
&& + D_k D_{(i_l}\cdots D_{i_1)}\hat \xi^*_i \hat \xi^k,\label{b6}
\eea
with $\hat F^i=(1/2) F^i_{jk}\hat \xi^j\hat \xi^k$ and
$\hat T^i=(1/2) T^i_{jk}\hat \xi^j\hat \xi^k$.

In order to compute the BRST cohomology, we follow closely the
reasoning of \cite{BBH3}, section 7. Apart from global considerations,
the generators $U^t,V^t$ belong to the contactible part of the algebra
and can be forgotten in the rest of the considerations.
For the remaining generators, one first decomposes the
cocycles, the coboundaries and the BRST differential according to the
number of $\hat \xi^i$'s. The BRST differential decomposes
into $s_\omega=s_0+s_1+s_2$. The part $s_0$ can be written as
$s_0=\delta+\gamma_{\cal G}$, where the Koszul-Tate differential
$\delta$ is
defined by the first lines of (\ref{b5}) and (\ref{b6}) alone and
$\gamma_{\cal G}$ is defined $\gamma_{\cal G}Y=-\delta_k Y \hat C^k$
for $Y\in \hat\xi^i, {\cal T}^r, {\cal T}^*_s$ and $\gamma_{\cal
G}C^i={1\over 2}\epsilon^i_{\ jk}\hat C^j\hat C^k$. The part $s_1$ is
given  by $s_1 {\cal T}^r=D_k {\cal T}^r\hat \xi^k$,
$s_1 {\cal T}^*_s=D_k {\cal T}^*_s\hat \xi^k$ and
$s_1\hat \xi^i=-\hat T^i$.
 Finally, the part $s_2$ acts as $s_2 \hat C^i=-\hat F^i$, where
the $s_i$'s, $i=0,1,2$, vanish on the other generators.

The anticommutation relations between the $s_i$'s are the same as
those in \cite{BBH3}, where for the proof of Eq. (7.23) of
\cite{BBH3},
one has to use the Jacobi identities (\ref{b1}) and (\ref{b2}).
Lemma 1 of \cite{BBH3} then stays true with $\omega_i(\hat C)$
either given by a constant, or by
$-{2\over 3}{\rm Str} \hat C^3$, where Str denotes the symmetrized
trace of the matrices.

We will now analyze equations (7.29), (7.30) of \cite{BBH3} directly
and not follow entirely Appendix E of that paper,
because our theory does not fulfill the normality assumption
needed in that approach.

By using the decomposition of the
variables ${\cal T}^r$ defined in section \ref{la}, we can first
assume because of (7.30) of \cite{BBH3}, that the invariant
$\alpha^i_l$ only
depends on the ${\cal T}^{\prime r}$'s and the $\hat \xi$'s.
Because $s_1$ commutes with the operator counting the generators
(\ref{4.3}), we then can take the equalities (7.29),(7.30)
of \cite{BBH3} to be
strong equalities and assume that $\beta^i_{l-1}$ is invariant and
also only depends on
the ${\cal T}^{\prime r}$'s and the $\hat \xi$'s.
The equations then become $s\alpha^i_l=0$ and $\alpha^i_l=
s\beta^i_{l-1}$. From the descent equations argument of section 6 in
\cite{BBH3}, one concludes that, if $l<3$,
$\alpha^i_l=s\gamma^i$ for some $\gamma^i$ depending on ${\cal T}^r,
\hat \xi, \hat C^i$. We can now use Appendix E
of \cite{BBH3} starting from equation (E.4). It is at this stage,
because we use Appendix C of \cite{BBH3} that we have to assume that
our local functions depend polynomially on the ${\cal T}^r$ variables.
Because we are in three dimensions and there are no abelian factors,
we conclude that equation (E.21) of \cite{BBH3} holds with $P(\hat
F)=0=q^*=G^*$ and a dependence on ${\cal T}^\prime$ rather than
${\cal T}$. The same is true for equation (7.34) of \cite{BBH3}.

Let $\hat \theta={1\over 3!}\epsilon_{ijk}\hat \xi^i\hat \xi^j
\hat \xi^k$.
Let $q=-{2\over 3}\ {\rm Str} \hat C^3 +{\rm Str} \hat C\hat F $.

The final result is that the BRST cohomology $H^*(s_\omega)$
of the model is described by
\bea
\hat \theta (L_1({\cal T}^\prime)+L_2({\cal T}^\prime)\
{\rm Str} \hat C^3+rq+s\beta,\ r\in {\bf R}.
\eea
So all the BRST cohomology is concentrated in ghost number $3$ and
$6$. The local BRST cohomology in space
$H^{*,*}(s_\omega|d)$ is obtained from $H^*(s_\omega)$ by replacing
$\eta^a$ by $\eta^a+dx^a$ \cite{B,BBH3}. Hence these
groups can be described by
\begin{eqnarray}
&& H^{0,3}(s_\omega|d): d^3x\ \tilde e\ L_1({\cal T}^\prime)+r^\prime
{\rm Str} (AF-{1\over 3}A^3),
\\
&& H^{3,3}(s_\omega|d): d^3x\ \tilde e\ L_2({\cal T}^\prime)\
{\rm Str}C^3,
\\
&& H^{1,2}(s_\omega|d): dx^a\wedge dx^b{\p \over\p \eta^a\eta^b}\ q,
\\
&& H^{2,1}(s_\omega|d): dx^a{\p \over\p \eta^a}\ q
\\
&& H^{3,0}(s_\omega|d): q,
\end{eqnarray}
where the solutions involving $L_1({\cal T}^\prime),L_2({\cal
T}^\prime)$ are trivial if they are given by $D_iM^i({\cal T}^\prime)$.
Note that there is no cohomology in ghost number $-1$ and form degree
3 which, by using the isomorphism of this group with the
non trivial conservation laws of the constraint surface \cite{BBH},
excludes the latter.

It then follows from the relation between the local Hamiltonian BRST
cohomology groups and the local Lagrangian BRST cohomology groups
\cite{JMP} and from the fact that the first class Hamiltonian is zero,
that the Lagrangian local BRST cohomology groups can be entirely
described by the local Hamiltonian BRST cohomology groups and in
particular, $H^{-1,4}(s|d)$ in spacetime, with $s$ the BRST
differential associated to the minimal solution of the
Batalin-Vilkovisky master equation \cite{BV}
for the Husain-Kucha\v r model, is isomorphic to
$H^{0,3}(s_\omega|d)$. Using again the reasoning of \cite{BBH}, this
last space also describes the local conservation laws of the
latter model.

\end{document}